# Bi-modal photothermal/optical microscopy for complementary bio-imaging with high resolution and contrast

Sergiy Litvinenko[1,3], Pavlo Lishchuk[2,3,a], Vladimir Lysenko[4] and Mykola Isaiev[5]

[1] *Institute of High Technologies, Taras Shevchenko National University of Kyiv, 64/13, Volodymyrska street, Kyiv, Ukraine, 01601*

[2] *Faculty of Physics, Taras Shevchenko National University of Kyiv, 64 Volodymyrska street, Kyiv, Ukraine, 01601*

[3] *Science Park Kyiv Taras Shevchenko University, 60, Volodymyrska Street, 01033 Kyiv, Ukraine;*

[4] *Light Matter Institute, UMR-5306, Claude Bernard University of Lyon/CNRS, Université de Lyon 69622 Villeurbanne cedex, France*

[5] *Université de Lorraine, CNRS, LEMTA, Nancy F-54000, France*

[a]*E-mail:* pavel.lishchuk@univ.kiev.ua

## Abstract

In the paper, combined bimodal microscopic visualization of biological objects based on reflection-mode optical and photoacoustic measurements was presented. Gas-microphone configuration was chosen for the registration of photothermal response. Precise positioning of the scanning laser beam with modulated intensity was performed employing acousto-optic deflectors. Photoacoustic images are shown to give complementary details to the optical images obtained in the reflected light. Specifically, the photoacoustic imaging mode allows much better visualization of the features with enhanced heat localization due to the reduced heat outflow. For example, the application of the photoacoustic imaging mode was especially successful to visualize Drosophila fly's micro-hairs. Furthermore, the photoacoustic image quality was shown to be adjusted through modulation frequency.

## 1. Introduction

The photoacoustic (PA) effect is a generation of acoustic waves due to absorption of electromagnetic radiation by a media. This phenomenon was observed for the first time by A.G. Bell 140 years ago [1], and it was started to be widely used after the work of Rosencwaig and Gersho [2], where they developed a quantitative theory of PA response formation in solids based on the thermal wave formalism. From the 90s, the PA effect was considered a promising tool for various biomedical applications. Specifically, the main advantage of the approach is the possibility to use the same laser source as for the excitation of PA response for diagnostics of diseases as well for its treatment [3,4].

Since that time, various methods for material diagnostics, based on the PA effect, have been developed qualitatively and quantitively. Among the PA-based techniques, we should pay attention to methods that can be used directly for measuring the thermophysical, mechanical, and optical properties of materials [5–12]. The various photoacoustic techniques have been developed to provide innovative solutions for spectroscopy [13,14], microscopy [3,15,16] and tomography [3,17–25]. Specifically, the latter is widely used in fundamental life science and medical practice for the visualization over a wide range of length scales covering from organelles and single cells [17,26,27] to tissues and organs [20–23,25,28].

The use of PA-based approaches in microscopy gives the possibility to overcome the optical limit, for instance, by localization of the heat due to the use of short pulses [29]. Thus, PA imaging techniques could be used noninvasively "in vivo" and provide useful information in anatomy imaging [30–32], neurology [33,34], cardiology [35,36], dermatology [3,19,28], oncology [3,21,28,37], and other preclinical and clinical applications for both humans and animals. However, utilization of the pulsed laser radiation with the power required for the characterization of the object situationally can lead to its significant overheating.

Therefore, the frequency-domain PA microscopy with respectively low intensity of radiation is preferable in numerous applications [5,10,38]. It should be noted that in this case, the level of signal can be small formed based on the deformations of a piezoelectric transducer in traditional configurations. From this point of view, the use of the gas-microphone (GM) PA approach is promising. As an example, the GM PA method has already been applied for the imaging of blood cells, and it was revealed its good potential in biology [39]. However, the authors used a resonance PA cell, and its application for the object with complex three-dimensional morphology is problematic.

In this article, we present the results of PA microscopy in GM configuration. For the modulation of the laser radiation and its positioning an electro-optical modulating system was used. A front configuration GM PA cell [40] with flat amplitude-frequency characteristic (out of resonance) in considered frequency range was used. It allowed us to obtain PA images for several frequencies of laser modulation and to visualize the impact of the modulation frequency on the contrast of the image. A fruit fly (drosophila melanogaster) was used as bio-object for imaging with the PA microscope.

## 2. Experimental details

Fig. 1(a) presents a schematic sketch view of the experimental setup, which consists of the three main blocks: (i) system for formation and deviation of a laser beam (previously reported in [41,42]); (ii) system for signal formation (PA cell coupled with a photodetector of the reflected light, schematically depicted in Fig. 1(b)); and (iii) system for signal detection and image formation.

*System for formation and deviation of a laser beam (i).* The block (i) includes an operational power source, laser source ($\lambda$ = 532 nm, with output optical power on a scanning area ~ 5 mW), acousto-optic deflectors coupled with the optical system (in-build on equipment) for focusing of the laser beam on a scanned area with the ability to achieve down to µm-scale lateral resolution (both the spot and the minimal scanning step size are 10 µm), a backing stage for positioning and fixation of the device on the microscope. A significant advantage of the use of the acousto-optic deflectors to control the deviation of the laser beam is the possibility of prevention of the motion of a PA cell during the scanning process. Beam focusing was carried out by placing the plane of the object in the focal plane of the optical system using the adjustment probe method in the reflected light mode. The lateral resolution analysis of the system was carried out with the calibration on optically transparent plates that contain absorbing rings with different scales.

*System for signal formation (ii).* The non-resonant GM PA cell in a front configuration [7,9,40] was used for signal formation by the microscope. The material of the PA cell is aluminium, the maximum scanning area is 5x5 mm$^2$. Calibration analysis of samples surface with homogeneous thermophysical and optical properties indicated no angular dependence of light incidence within the scanned area for both configurations (optical and photoacoustic). The inner working space of the PA cell was isolated from the external environment and a studied object can be scanned through the optically transparent glass window. The photothermally induced pressure oscillations coming from the object localized inside the PA cell were recorded by an electret microphone PX-53 built-in into the cell. An air channel between the laser illumination area and the microphone membrane has, respectively, a length and diameter of 2 mm. The main purpose of the Si-based photodiode installed outside of the PA cell was to fix the amount of light reflected from the studied object and emitted from the cell through the optically transparent glass.

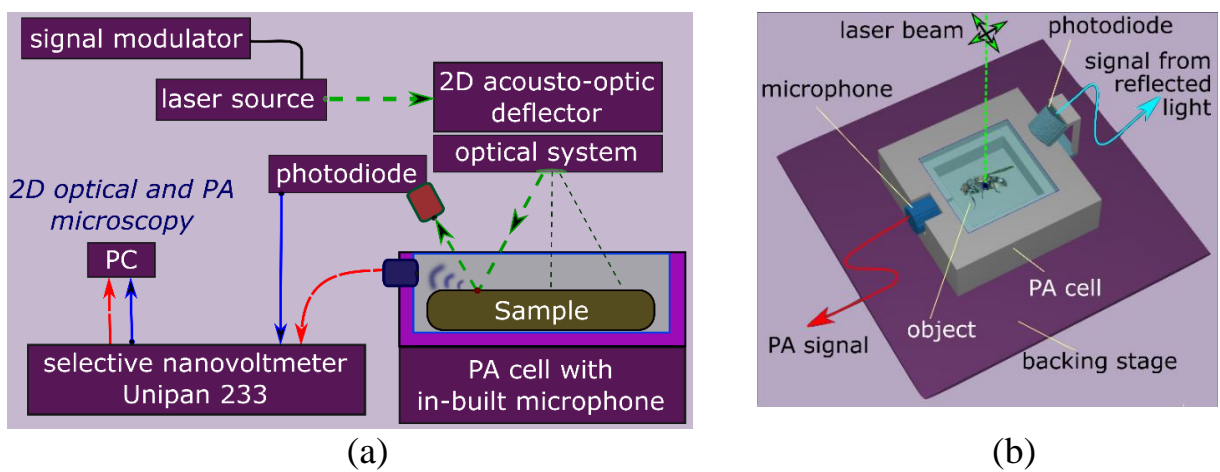

(a)      (b)

**Fig. 1**. Experimental set-up for the bimodal photoacoustic/optical reflection microscopy (a), the used PA cell (b)

*System for signal detection and image formation (iii).* Amplitudes of the signals issued from both the photodiode and the microphone were recorded separately at each laser beam position by selective nanovoltmeters (Unipan type 233), with subsequent registration on a PC. Photoacoustic images and images obtained in reflected light were formed by amplitudes of the signal in each photo-excited point. Thus, for both methods, the image reconstruction procedure is based on the following of the beam path.

*Test scans in bimodal PA/optical reflection imaging.* Firstly, to test the bi-modal imaging configuration, PA and optical reflection images of a polyvinyl chloride black tape deposited on the surface of an aluminium substrate put into the PA cell were obtained. The tape was angled at about 45º on the right top corner of a scanning area, as shown in Fig. 2. Scanning of the tape was carried out sequentially along the lines giving a 2D image of the scanned area (see Fig. 2 (a)).

A typical dependence of the amplitude of the optical and PA signals within one scanning line is shown in Fig. 2 (a). The graph shows that the amplitude of the signal received from the reflected light is significantly higher for the aluminium substrate than for the region where the tape is fixed. It is obvious because aluminium has a higher reflectivity for a given irradiation wavelength. Additionally, some microscopic surface roughness of the tape (shown in the figure Fig. 2 (b) as light spots at equidistance on the dark area) and aluminium substrate irregularities can be also visible (see the bright area of Fig. 2 (b) where the polishing of the aluminum surface in the form of stripes is clearly visible).

Regarding the PA imaging, as one can see in Fig. 2 (c), the PA signal from the black tape has a higher level compared to the response from Al substrate. Contrary to optical imaging based on the reflected light, the amplitude of the PA signal depends on the amount of light absorbed by the sample. Moreover, one can observe an increase of the signal on the edge of the tape (indicated by a black arrow in Fig. 2(a)). The latter arises because of the limitation of the heat outflow from the edge region forming an interface with air having a much lower thermal conductivity value. Thus, as we can see, the PA response gives useful complementary information to the reflection-mode optical measurements.

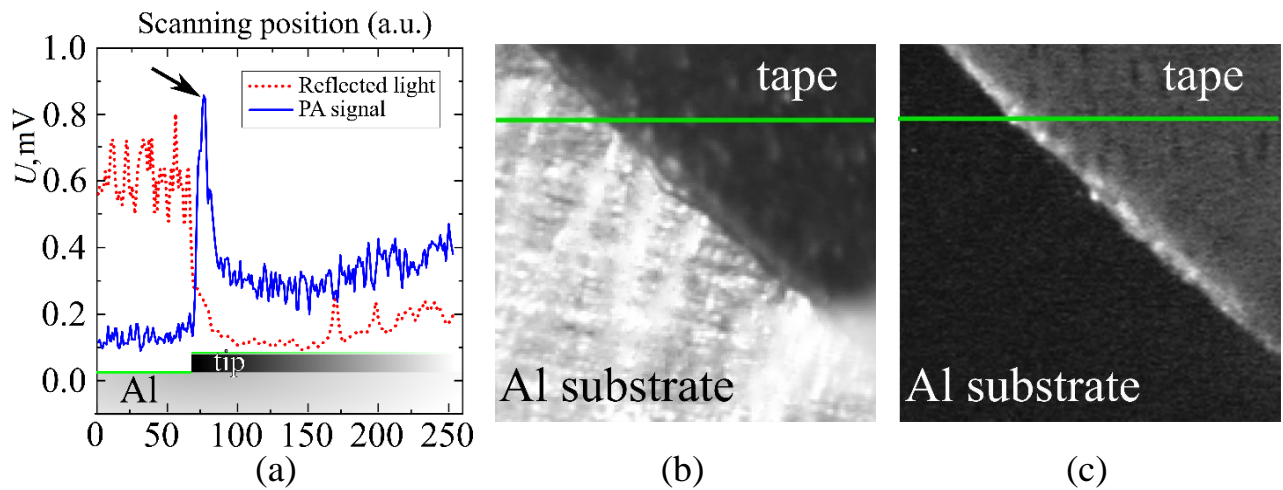

**Fig. 2**. a) Dependence of the amplitudes of the reflected and PA signals from polyvinyl chloride black tape and aluminium substrate for the line scan mode; b) reflection-mode optical image; c) PA image.

### 3. Results and discussions

A small fruit fly, Drosophila melanogaster (see Fig. 3(a)), was used as a studied object for demonstration of the performances of such bimodal 2D imaging microscopy. The size of an adult Drosophila fly normally does not exceed a few millimetres. Fig. 3(b) presents the image of the fly in the reflected light, and it correlates very well with the optical image in Fig. 3(a).

Fig. 3(c) and 3(d) show the images obtained with the use of the PA technique at different modulation frequencies (75 Hz and 735 Hz respectively). The PA signal levels were up to 1 mV at 75 Hz and up to 100 µV at 735 Hz, respectively. The maximal PA signals correspond to bright colour on the PA images. First, it should be noted that the PA mode gives better visualization quality of small features from which the heat outflow is difficult. Thus, when the optical microscopy is complicated, for example, through the poor reflectivity of the studied object, PA imaging mode becomes an attractive alternative, because the PA signal depends not only on the object's optical properties but also on its thermophysical properties in the scanning area. For example, hairs on the surface of an insect that during the fly`s life are served as tiny 'feelers' and heat regulators can be more clearly seen in the images obtained by the PA approach.

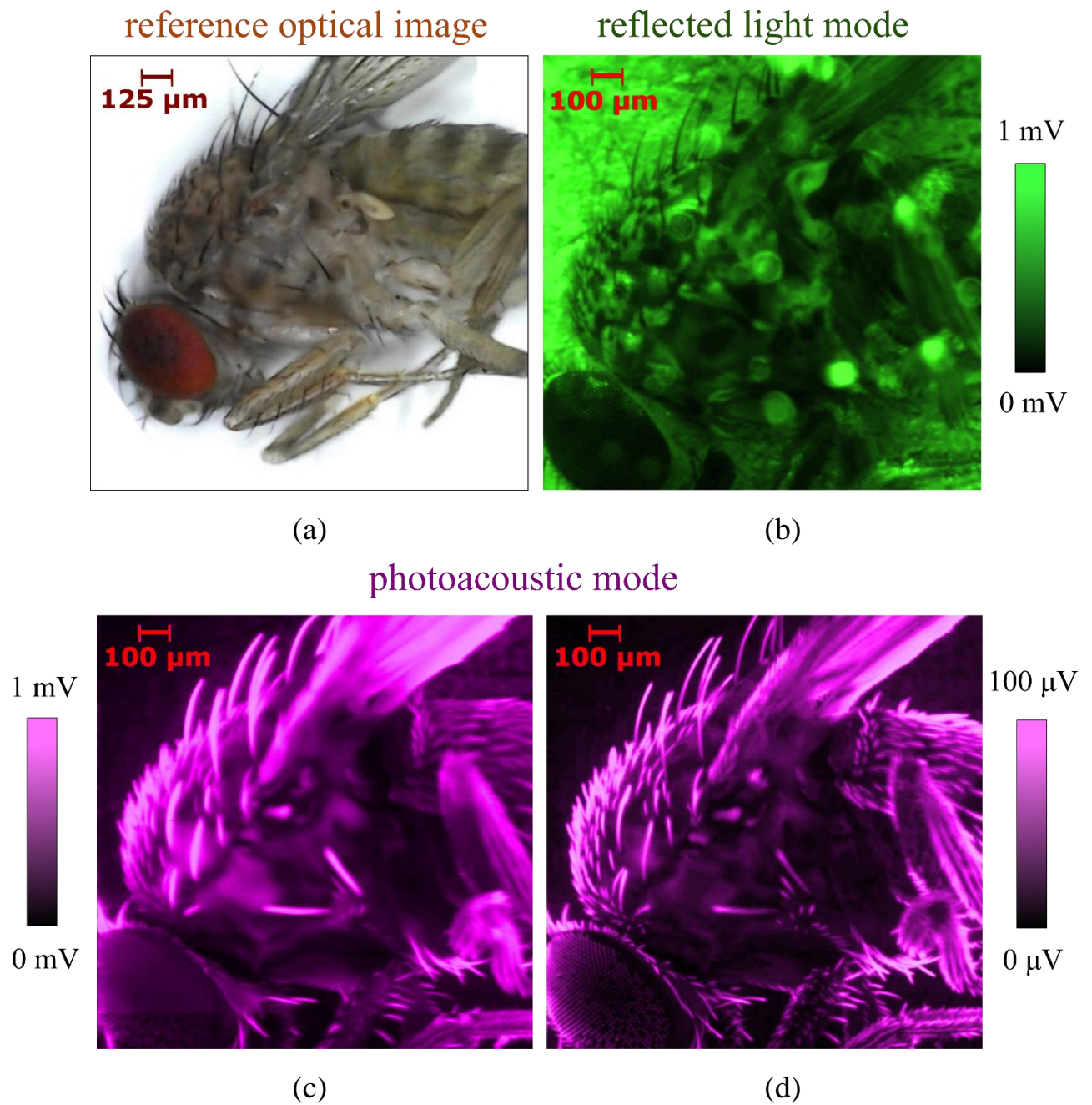

**Fig. 3.** Images of Drosophila fly: a) obtained as a reference picture by a portable optical microscopy for general visualization of the object; b) in the reflected light mode provided by our experimental set-up; in the PA mode under 75 Hz (c) and 735 Hz (d) modulation frequencies. Scale grades of the response signal are applicable to figures (b) – (d).

A comparison of the PA images obtained under different modulation frequencies shows, that the increase of the frequency values leads to an increase in image sharpness. The latter may be explained by a frequency-dependent size of the heat localization region:

$$l \sim \sqrt{\frac{c\rho\omega}{\kappa}},$$

where $l$ is the linear size of the region, $c$ and $\rho$ are, respectively, specific heat capacities and density of the media, $\omega$ is the cyclical frequency of modulation, $\kappa$ is the thermal conductivity of the media. In the case of a complex heterogeneous media, one can consider all the parameters mentioned above as effective ones. Thus, there is the possibility to adjust the image quality by modulation frequency.

Additionally, an increase of the modulation frequency leads to a decrease in the relaxation time to reach a study state regime. Therefore, it gives the possibility to significantly optimize a scan rate. On another hand, the increase of the modulation frequency induces a decrease of the PA signal level, thus a satisfying balance should be found.

Fig. 4 presents images of the Drosophila fly close to the eye region. One can see that the reflection-mode optical imaging gives better visualization of the body of the insect, while the image obtained in the PA mode microscopy gives better visualizations of the regions where photo-induced thermal energy is strongly localized. Specifically, it is known that the eyes of a fly are composed of separate structural units called ommatidia. The outer lenses of ommatidia are closely spaced among themselves and, when viewed from the surface, look like hexagons. Between each ommatidia unit, fruit flies possess evenly spaced hair-like structures called inter ommatidia bristles [43,44]. These ocular hairs reduce airflow at the eye surface by up to 90% crucially reducing particle deposition and damage probability of the eye during the insects fly. At the same time, these micro-hairs minimally prevent the incoming light on the eye thereby, provide good vision [45]. The outer lenses of ommatidia can be clearly seen in an optical microscope, while ommatidia bristles can be observed qualitatively only in images obtained by a scanning electron microscope. Nevertheless, the 1D limitation of the ocular hairs structure complicates heat dissipation when it is excited by laser irradiation. This important factor allows to clearly see the hairs on the eye of the fly in PA mode. Thus, the image of ommatidia hexagons, which was obtained in the reflected light scanning mode, is complemented by the image of ommatidia hairs from the PA mode, creating the integrity of information about the structure of the fly's eye (see the markers 2 on Fig.4).

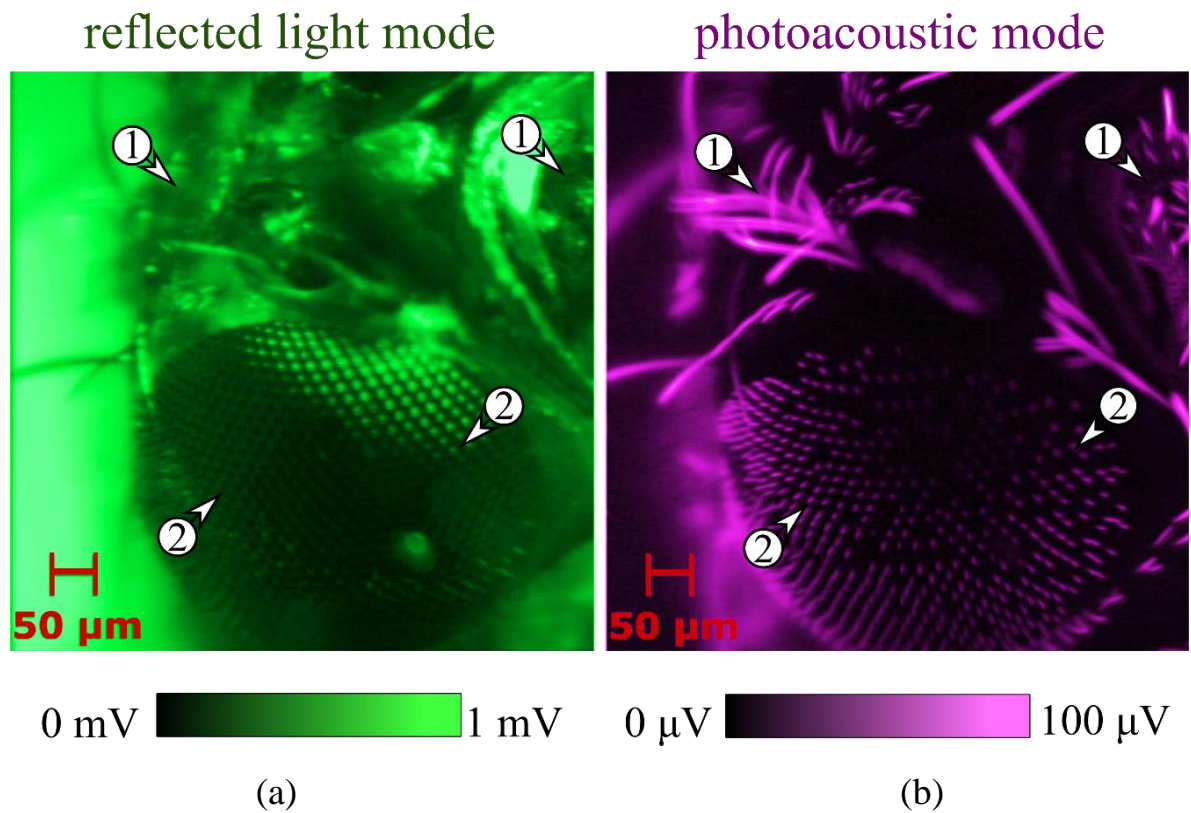

**Fig 4.** Images of Drosophila fly eye obtained in the reflected light mode (a) and in PA mode under 340 Hz modulation frequency (b). The markers in the images indicate the examples of characteristic differences in the visualization of the object that is scanned bi-modally. Marker (1) indicates the difference in signal from the body of the hair-covered insect; marker (2) indicates the discrepancy in the visualization of the eye of the object.

### 4. Conclusion

In this work, the application of the photoacoustic technique for microscopy of biological objects is reported. A photoacoustic method with gas-microphone registration was chosen for detections of laser-induced thermal perturbations. Scanning of the modulated laser beam was performed with the use of acousto-optic deflectors. The photothermal microscopy of a fruit fly was used to demonstrate the contrast and resolution levels of the obtained images. The dependence of these parameters on the modulation frequency is demonstrated. Photoacoustic images are shown to give complementary details to the optical images obtained in the reflected light. In particular, it was revealed that contrary to the optical image, the PA scanning mode allows to clearly see the hairs on the body and legs of the fruit fly, as well as ommatidia bristles, which are the structural units of the structure of the insect's eye. Thus,-bimodal microscopy based on both visualization approaches can be applied in different fields where the use of a single imaging mode is not satisfying.

## Acknowledgement

This research work was carried out in frames of the CARTHER project (proposal #690945) of Marie Skłodowska-Curie Research and Innovation Staff Exchange program. The publication contains the results obtained in the frames of the research work "Features of photothermal and photoacoustic processes in low-dimensional silicon-based semiconductor systems" (Ministry of Education and Science of Ukraine, state registration number 0118U000242).

## References

1. S. Manohar and D. Razansky, Adv. Opt. Photonics **8**, 586 (2016).

2. A. Rosencwaig and A. Gersho, J. Appl. Phys. **47**, 64 (1976).

3. S. Jeon, J. Kim, D. Lee, J. W. Baik, and C. Kim, Photoacoustics **15**, 100141 (2019).

4. W. W. Liu and P. C. Li, J. Biomed. Sci. **27**, 3 (2020).

5. M. Isaiev, P. J. Newby, B. Canut, A. Tytarenko, P. Lishchuk, D. Andrusenko, S. Gomès, J.-M. Bluet, L. G. Fréchette, V. Lysenko, and R. Burbelo, Mater. Lett. **128**, 71 (2014).

6. C. F. Ramirez-Gutierrez, J. D. Castaño-Yepes, and M. E. Rodriguez-García, J. Appl. Phys. **121**, 025103 (2017).

7. P. Lishchuk, A. Dekret, A. Pastushenko, A. Kuzmich, R. Burbelo, A. Belarouci, V.

Lysenko, and M. Isaiev, Int. J. Therm. Sci. **134**, 317 (2018).

8. K. Dubyk, A. Pastushenko, T. Nychyporuk, R. Burbelo, M. Isaiev, and V. Lysenko, J. Phys. Chem. Solids **126**, 267 (2019).

9. P. Lishchuk, M. Isaiev, L. Osminkina, R. Burbelo, T. Nychyporuk, and V. Timoshenko, Phys. E Low-Dimensional Syst. Nanostructures **107**, 131 (2019).

10. K. Dubyk, L. Chepela, P. Lishchuk, A. Belarouci, D. Lacroix, and M. Isaiev, Appl. Phys. Lett. **115**, 021902 (2019).

11. C. F. Ramirez-Gutierrez, H. D. Martinez-Hernandez, I. A. Lujan-Cabrera, and M. E. Rodriguez-García, Sci. Rep. **9**, 1 (2019).

12. K. Dubyk, T. Nychyporuk, V. Lysenko, K. Termentzidis, G. Castanet, F. Lemoine, D. Lacroix, and M. Isaiev, J. Appl. Phys. (2020).

13. L. A. Skvortsov, Quantum Electron. **43**, 1 (2013).

14. P. Patimisco, G. Scamarcio, F. K. Tittel, and V. Spagnolo, Sensors (Switzerland) **14**, 6165 (2014).

15. K. L. Muratikov, A. L. Glazov, D. N. Rose, and J. E. Dumar, J. Appl. Phys. **88**, 2948 (2000).

16. J. Kim, J. Y. Kim, S. Jeon, J. W. Baik, S. H. Cho, and C. Kim, Light Sci. Appl. **8**, (2019).

17. L. V. Wang and S. Hu, Science (80-. ). **335**, 1458 (2012).

18. J. Yao and L. V. Wang, Laser Photonics Rev. **7**, 758 (2013).

19. L. Li, Pathobiol. Hum. Dis. A Dyn. Encycl. Dis. Mech. 3912 (2014).

20. S. Park, C. Lee, J. Kim, and C. Kim, Biomed. Eng. Lett. **4**, 213 (2014).

21. J. Yao, A. A. Kaberniuk, L. Li, D. M. Shcherbakova, R. Zhang, L. Wang, G. Li, V. V. Verkhusha, and L. V. Wang, Nat. Methods **13**, 67 (2015).

22. Y. Zhou, J. Yao, and L. V. Wang, J. Biomed. Opt. **21**, 061007 (2016).

23. K. Park, J. Y. Kim, C. Lee, S. Jeon, G. Lim, and C. Kim, Sci. Rep. **7**, 1 (2017).

24. A. A. Plumb, N. T. Huynh, J. Guggenheim, E. Zhang, and P. Beard, Eur. Radiol. **28**, 1037 (2018).

25. E. Maneas, R. Aughwane, N. Huynh, W. Xia, R. Ansari, M. Kuniyil Ajith Singh, J. C. Hutchinson, N. J. Sebire, O. J. Arthurs, J. Deprest, S. Ourselin, P. C. Beard, A. Melbourne, T. Vercauteren, A. L. David, and A. E. Desjardins, J. Biophotonics **13**, 2 (2020).

26. E. M. Strohm, E. S. L. Berndl, and M. C. Kolios, Photoacoustics **1**, 49 (2013).

27. E. M. Strohm, M. J. Moore, and M. C. Kolios, Photoacoustics **4**, 36 (2016).

28. E. Vienneau, T. Vu, and J. Yao, Imaging Technol. Transdermal Deliv. Ski. Disord.

411 (2019).

29. C. Zhang, K. Maslov, and L. V. Wang, Opt. Lett. **35**, 3195 (2010).

30. K. Maslov, G. Ku, and L. V. Wang, Photons Plus Ultrasound Imaging Sens. 2010 **7564**, 75640W (2010).

31. L. Xi, L. Zhou, and H. Jiang, Appl. Phys. Lett. **101**, 1 (2012).

32. Z. Xie, S.-L. Chen, T. Ling, L. J. Guo, P. L. Carson, and X. Wang, Opt. Express **19**, 9027 (2011).

33. L. De Liao, M. L. Li, H. Y. Lai, Y. Y. I. Shih, Y. C. Lo, S. Tsang, P. C. P. Chao, C. T. Lin, F. S. Jaw, and Y. Y. Chen, Neuroimage **52**, 562 (2010).

34. J. Yao, J. Xia, K. I. Maslov, M. Nasiriavanaki, V. Tsytsarev, A. V. Demchenko, and L. V. Wang, Neuroimage **64**, 257 (2013).

35. A. Taruttis, E. Herzog, D. Razansky, and V. Ntziachristos, Opt. Express **18**, 19592 (2010).

36. L. Wang, K. Maslov, W. Xing, A. Garcia-Uribe, and L. V. Wang, J. Biomed. Opt. **17**, 1060071 (2012).

37. S. Zackrisson, S. M. W. Y. Van De Ven, S. S. Gambhir, S. Zackrisson, S. M. W. Y. Van De Ven, and S. S. Gambhir, Cancer Res. **74**, 979 (2014).

38. H. Tang, Z. Tang, Y. Wu, Q. Cai, L. Wu, and Y. Chi, Opt. Lett. **38**, 1503 (2013).

39. K. Sathiyamoorthy, E. M. Strohm, and M. C. Kolios, J. Biomed. Opt. **22**, 046001 (2017).

40. B. Abad, Renew. Sustain. Energy Rev. **76**, 1348 (2017).

41. S. Litvinenko, L. Ilchenko, A. Kaminski, S. Kolenov, A. Laugier, E. Smirnov, V. Strikha, and V. Skryshevsky, Mater. Sci. Eng. B Solid-State Mater. Adv. Technol. **71**, 238 (2000).

42. S. V. Litvinenko, A. V. Kozinetz, and V. A. Skryshevsky, Sensors Actuators, A Phys. **224**, 30 (2015).

43. R. Paul, J. S. Grewal, and V. C. Kapoor, in *Fruit Flies* (Springer New York, New York, NY, 1993), pp. 29–30.

44. A. Garg, A. Srivastava, M. M. Davis, S. L. O'Keefe, L. Chow, and J. B. Bell, Genetics **175**, 659 (2007).

45. G. J. Amador, F. Durand, W. Mao, S. Pusulri, H. Takahashi, V.-T. Nguyen, I. Shimoyama, A. Alexeev, and D. L. Hu, Eur. Phys. J. Spec. Top. **224**, 3361 (2015).